\documentclass[12pt]{iopart}
\usepackage{graphicx}
\usepackage{bm}
\usepackage{iopams}
\usepackage{color}

\usepackage[breaklinks=true,colorlinks=true,linkcolor=blue,urlcolor=blue,citecolor=blue]{hyperref}
\usepackage[normalem]{ulem}

\definecolor{grey}{rgb}{.6,.6,.6}

\begin{document}

\title{Quantum turnstile operation of single-molecule magnets}
\author{V. Moldoveanu}
\address{National Institute of Materials Physics, PO Box MG-7, Bucharest-Magurele,Romania}
\author{I. V. Dinu}
\address{National Institute of Materials Physics, PO Box MG-7, Bucharest-Magurele,Romania}
\author{B. Tanatar}
\address{Department of Physics, Bilkent University, Bilkent, 06800 Ankara, Turkey}
\author{C. P. Moca$^{1,2}$}
\address{$^1$ BME-MTA Exotic Quantum Phase Group, Institute of Physics,\\
Budapest University of Technology and Economics, H-1521 Budapest, Hungary}
\address{$^2$ Department of Physics, University of Oradea, 410087, Oradea, Romania}

\begin{abstract}
The time-dependent transport through single-molecule magnets coupled to magnetic or non-magnetic electrodes
is studied in the framework of the generalized master equation method.
We investigate the transient regime induced by the periodic switching of the source and drain contacts.
If the electrodes have opposite magnetizations the quantum turnstile operation allows
the stepwise writing of intermediate excited states. In turn, the transient currents provide a way to read these states.
Within our approach we take into account both the  uniaxial and transverse anisotropy.
The latter may induce additional quantum tunneling processes which affect the  efficiency of the proposed
read-and-write scheme. An equally weighted mixture of molecular spin states can be prepared if one of the electrodes is ferromagnetic.

\end{abstract}


\maketitle

\section{Introduction}

The single-molecule magnets (SMMs) are foreseen as building blocks of organic spintronic devices \cite{Bogani,Sanvito}.
Such systems generally behave as magnetic cores with a large localized spin and display slow relaxation of
magnetization at low temperature mostly due to presence of the anisotropy-induced magnetic barrier \cite{MolecularNanomagnets}.
Similar to quantum dot physics, two-terminal steady-state transport measurements performed on SMMs revealed charging
effects such as Coulomb blockade, sequential tunneling or negative differential conductance \cite{Heersche,Jo}.
In the spin sector, Kondo related features were observed experimentally \cite{Komeda,Otte,Loth,Parks}
and investigated theoretically \cite{MWB,Hurley}.
The exchange interaction between the local molecular moment and the delocalized spins
tunneling through the molecular orbital might be exploited to control the quantum
state of the local moment, i.e. to 'write' and 'read' its quantum state \cite{Zyazin}.

On the experimental side, various techniques \cite{Song} are currently  used to attach the orbitals (ligands)
surrounding the molecular magnetic core to the source and drain probes. Unlike standard transport setups used in quantum dot devices,
molecular electronics requires more careful handling of the contact regions. The difficult task of isolating a
single molecule between source and drain electrodes is nowadays realized by using more advanced methods as electromigration,
mechanically controlled break junctions \cite{Martin} or spin polarized STM
\cite{Wiesendanger}. Recently several groups pushed even further these techniques and reported controlled time-dependent
transport measurements for such SMMs when the contacts were switched on and off by varying the substrate-STM
tip spacing \cite{Kumar,Sotthewes} or by bending break junctions \cite{Ballmann}, and transient currents arising when a molecular tail couples to an STM tip have been recorded \cite{Kockmann}. 
SMMs have also been integrated into carbon nanotube transistors to serve
as detectors for the nanomechanical motion due to the strong spin-phonon coupling \cite{Urdampilleta,Ganzhorn}. 
In two cornerstone experiments Vincent {\it et al.} \cite{Vincent} and Thiele {\it et al.} \cite{Thiele} detected 
the nuclear spin of a single ${\rm Tb^{3+}}$ ion embedded in a SMM together with the Rabi oscillations.

These promising experiments motivated us to investigate the transient transport properties of SMMs,
with special emphasis on the regime when the couplings to the source/drain electrodes
are switched on and off periodically. The transport regime we are interested in is similar to the so called turnstile pumping
setup which represents a long-standing \cite{Kouwenhoven} asset of pumping or pump-and-probe experiments with quantum dots
\cite{Giblin}. Along this periodic pumping, the source and drain tunneling barriers open and close consecutively,
such that a single electron is transmitted across the sample; details on the turnstile operation will be presented in Section 3.
To our best knowledge the transient and the turnstile regimes have not been theoretically investigated so far
in the context of transport across SMMs.

On the theoretical side the magnetic interactions in SMMs are described by effective giant spin Hamiltonians
\cite{MolecularNanomagnets}, mostly because of the large value of the localized magnetic moment.
Using this description, in Ref. \cite{Timm2006} the
authors investigated the role of relaxation on inelastic charge and spin transport across
a SMM weakly coupled to metallic gates.
In Ref.~\cite{MB} transport across a SMM coupled to two ferromagnetic leads was investigated and it was found that
the spin current across the SMM can reverse the localized spin if the leads
 have opposite spin polarizations.
Memristive \cite{Timm2012} and thermoelectric  \cite{Wang} properties of SMMs were
also investigated.
All these studies convey a similar message: The current can induce magnetic switching
of the localized magnetic moment if the applied bias voltage exceeds the gap between the ground and excited states.

In the absence of transverse anisotropy the effect of quantum tunneling of magnetization (QTM) is negligible and the full
magnetic switching requires the {\it transient} occupation of all excited (intermediate) states with magnetic quantum numbers
in the range $[-S,S]$. When present, the QTM might leave its fingerprint on the transport properties at resonant
values of an applied magnetic field \cite{MB-Trans}.

A complementary approach to transport properties of SMMs relies on density functional theory (DFT) \cite{Renani,Barrazza}.
In the DFT approach the molecular structure and the contact regions are carefully taken into account, while the many-body
correlations within the SMM are accounted for within various approximations. A detailed {\it ab initio} Hubbard many-body model for
molecular magnets has been recently implemented \cite{Chiesa} and allows the calculation of magnetic interactions.

In the present work we investigate transient transport and turnstile pumping across
a SMM embedded between magnetic and non-magnetic electrodes. As we are interested
in the time dependent evolution of the currents and the accumulation of the
charge and spin on the SMM we rely our investigation on the generalized master equation
(GME) technique \cite{Moldo1,PRB-I}. Let us stress that, to capture the turnstile regime, one has to go beyond
the steady-state rate-equation approach.
We find that by setting the SMM in the quantum turnstile (QT) configuration
with ferromagnetic leads one can address two new issues which are relevant for the use of molecular states as magnetic qubits:
(i) the one-by-one all-electrical writing and reading of excited molecular
states with spin $S-1,..,-S+1$ ($S$ being the molecular spin of the initial ground state) and (ii)
the controlled preparation of statistical mixtures of such intermediate states. So far, the {\it stepwise} magnetic switching
protocols for excited molecular states that we propose here have not been investigated. In fact, previous studies
(see e.g. Refs. \cite{MB,Timm2006})
were focused only on investigating the full magnetic switching. The second issue was partially
addressed by Tejeda {\it et al.} \cite{Tejeda} some time ago. Their proposal concerns the preparation of equal weight
superposition of states (e.g. $|\psi\rangle =1/\sqrt{2}(|S\rangle+|S-1\rangle)$) using at least two molecular
clusters embedded in micro-SQUIDs.

The rest of the paper is organized as follows: In Section 2 we
present the theoretical framework by introducing the model Hamiltonian and
giving a summary of the GME method. Section 3 presents the main
results of our work while in Section 4 we give the conclusions.

\section{Theoretical framework}

\subsection{Model Hamiltonian}

The setup that we consider here consists of a SMM coupled to two external electrodes (see the sketch in Fig.\,1).
We investigate time dependent transport in the sequential tunneling
regime, in which the electrons tunnel one by one from the left (source) electrode to the unoccupied molecular
orbitals of SMM and then escape to the right (drain) electrode. In the present work
we neglect all other possible transport mechanisms, such as the cotunneling processes
which are responsible for the Kondo effect \cite{MWB}.

The model Hamiltonian contains several terms describing the SMM itself ($H_M$), the left ($H_L$) and
right ($H_R$) electrodes and the time-dependent tunneling part $H_T$:
\begin{equation}\label{Htotal}
H(t)=H_M+H_L+H_R+H_T(t).
\end{equation}
In general, SMMs are characterized by a large spin $S>1/2$.
They do also present transverse anisotropy and an easy axis of magnetization \cite{MolecularNanomagnets}.
Assuming that the relevant contribution to the transport comes only from the lowest unoccupied molecular orbital
(LUMO) the SMM can be modelled by an effective, minimal Hamiltonian \cite{Timm2006,MB} of the form:
\begin{eqnarray}\label{H_mol_1}
H_M&=&\varepsilon\,\hat n +U{\hat n}_{\uparrow}{\hat n}_{\downarrow}
-J\, {\bf\hat s}\cdot {\bf\hat S}-D{\hat S}_z^2+E({\hat S}_x^2-{\hat S}_y^2)-g\mu_B B{\hat S}_z^t.\label{H_mol}
\end{eqnarray}
In its simplest form the LUMO orbital consists of a single spinful interacting level with
energy $\varepsilon$, on-site Coulomb energy $U$ and occupation ${\hat n}={\hat n}_{\uparrow}+{\hat n}_{\downarrow}$,
and is coupled to the localized spin ${\bf\hat S}$ through an exchange interaction with a coupling strength $J$.
The fourth and the fifth terms in Eq.~(\ref{H_mol_1}) describe the easy-axis and transverse anisotropy with the
corresponding constants denoted by $D$ and $E$. For certain molecules \cite{Mannini} $E\ll D$ but
this is not always true, as $E$ can grow up to $D/3$ in other situations \cite{MolecularNanomagnets}.
In the presence of an external magnetic field $B$ pointing in the $z$-direction,
a Zeeman term is supplemented in Eq.~(\ref{H_mol_1}), with $g$ and $\mu_B$ being the
gyromagnetic factor and the Bohr magneton, respectively. In view of further discussions
we single out the transverse anisotropy term and write $H_M$ in Eq.~(\ref{H_mol_1}) as
\begin{equation}\label{eq:H_M_0}
H_M=H_M^0+E({\hat S}_x^2-{\hat S}_y^2).
\end{equation}
The reason behind this separation is that, contrary to $H_M$, $H_M^0$ has an extra abelian $U(1)$ symmetry generated
by the $z$-component of the total spin ${\hat S}_z^t={\hat S}_z+{\hat s}_z$. Consequently,
the eigenstates of $H_M^0$ can be organized according to the eigenvalues $m$ of ${\hat S}_z^t$ \cite{good}.
 On the other hand the transverse anisotropy term does not commute with ${\hat S}_z^t$
 and needs to be treated separately. We shall discuss in more detail the eigenstates of $H_M^0$ in
section ~\ref{sec:Eigenstates}.

The source and drain electrodes are modeled as spin-polarized one-dimensional discrete chains which in the momentum space
representation are described by the Hamiltonians:
\begin{equation}
H_{\alpha} = \sum_{\sigma}\int_{0}^{\pi}\; dq_{\alpha}\;  \varepsilon_{q_{\alpha}\sigma}\;
a^\dagger_{q_\alpha \sigma} a_{q_\alpha \sigma},  \;\;\;\;\; \alpha=\{L,R\}.
\label{eq:H_leads}
\end{equation}

\begin{figure}[tbhp!]
\includegraphics[angle=0,width=0.5\textwidth]{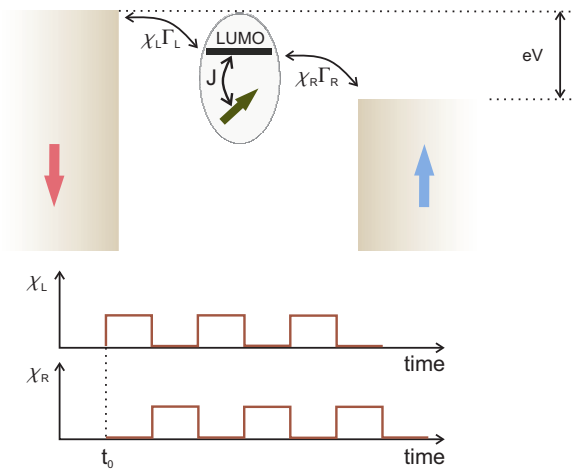}
\centering
\caption{(Color online) A sketch of the SMM coupled to source and drain electrodes via time-dependent tunneling barriers.
The tunneling amplitudes are controlled by the switching functions $\chi_L$ and $\chi_R$. The turnstile operation
consists in turning periodically on and off the contacts. Note that the left contact opens first, for the
charging sequence, while the right contact couples later for the discharge/depletion sequence. The chemical potentials of the
leads are chosen such that $\mu_L>\mu_R$; the bias $eV=\mu_L-\mu_R$. $t_0$ is some initial time. The leads are spin polarized in
the so called antiparallel (AP) configuration.}
\label{fig1}
\end{figure}

Both leads present an energy dispersion law of the form $\varepsilon_{q\sigma}=2\tau\cos q+\Delta_{\sigma}$,
 with $\tau$ the effective hopping between the nearest neighbor sites in the leads and $\Delta_{\sigma}$ a rigid-band
spin splitting that describes the polarization of the leads. In Eq.~(\ref{eq:H_leads}) $ a^\dagger_{q_\alpha\sigma}$ creates
an electron with momentum $q$ and spin $\sigma$ in the lead $\alpha = \{L, R\}$.

The last term in Eq.~(\ref{Htotal}) describes the hybridization of the SMM with the contacts
\begin{equation}
H_T(t)=\sum_{\alpha=L,R}\sum_{\sigma}\int_{0}^{\pi} dq_{\alpha} \chi_{\alpha}(t)(V^{\alpha}_{\sigma}a^{\dagger}_{\sigma}a_{q_{\alpha}\sigma}+h.c),
\end{equation}
where $a^{\dagger}_{\sigma}$ creates an electron with spin $\sigma$ on the LUMO orbital and $V^{\alpha}_{\sigma}$
 is the hopping amplitude of a tunneling process between the LUMO and the
 majority ($\sigma=+$) and minority ($\sigma=-$) electron states in the lead $\alpha$.
  The coupling of the SMM to the contacts in the case of collinear magnetic configuration,
 and in the absence of the switching protocol ($\chi_{\alpha}(t)=1$),
   is described by $\Gamma_{\sigma}^{\alpha} = 2\pi |V^{\alpha}_{\sigma}|^2\,\varrho_{\alpha\sigma} (0) $, where
 $\varrho_{\alpha\sigma}(0)$ is the spin density of states at the Fermi surface for electrons in lead $\alpha$.

In view of further investigations we allow tunable spin polarizations in the leads and define
$P^{\alpha}:=(\Gamma_{+}^{\alpha}-\Gamma_{-}^{\alpha})/(\Gamma_{+}^{\alpha}+\Gamma_{-}^{\alpha})$.

Note that the Hamiltonian $H_T(t)$ contains two time-dependent dimensionless functions $\chi_{\alpha}(t)$ ($\alpha=L,R$)
which simulate the switching of the contacts between the molecule and the leads. As we are interested in the turnstile
pumping it is enough  to consider them  simply as rectangular periodic pulses (see Fig.~\ref{fig1})

\subsection{Energy eigenstates}\label{sec:Eigenstates}

In this section we discuss the energy spectrum and the organization of the eigenstates
of the SMM Hamiltonian $H_M$ introduced in Eq.~(\ref{H_mol_1}). We shall start by discussing first the
spectrum of $H_{M}^0$. When $J=0$, the LUMO orbital gets decoupled from the local spin
and  the Hamiltonian $H_{M}^0$ has three U(1) symmetries generated by the local charge $Q$ accumulated on the LUMO
and by the $z$-components of the LUMO and local spins, $\hat s_z$ and $\hat S_z$. Consequently,
$\{ Q, \hat s_z, \hat S_z\}$ provide the quantum numbers according to which the multiplets of the
Hamiltonian are classified. Notice that this low symmetry classification is valid for finite magnetic fields.
When $B=0$, the Hamiltonian $H_{M}^0$ has a much higher symmetry, i.e. $U_{Q}(1)\times SU_{\hat \mathbf s}(2)\times SU_{\hat\mathbf S}(2)$ in the charge and spin sectors, but this situation shall not be discussed here, as we always assume a finite magnetic field.  In the case we consider here,
the classification of the states  is rather trivial and we can simply denote the eigenstates as follows:
$|0,0,S_z\rangle$, $|1,\uparrow,S_z\rangle$, $|1,\downarrow,S_z\rangle$, and $|2,0,S_z\rangle$,
with $S_z=-S,-S+1,..,S$. In the presence of Coulomb interaction the double occupied states
$|2,0,S_z\rangle$ have an energy of the order $\sim U$, which is the largest energy scale in the problem,
and in view of the discussion that follows, shall not contribute to transport. Therefore, to simplify
the notations, it is enough to relabel the states and keep track of the $\hat s_z$ and $\hat S_z$
quantum numbers. In this new notation we have $|0, 0,S_z\rangle\to |0,S_z\rangle$ and $|1, \sigma,S_z\rangle\to |\sigma,S_z\rangle$.

A finite exchange coupling $J$ breaks the three U(1) symmetries down to $U_Q(1)\times U_{\hat S_z^t}(1)$ generated by LUMO charge $Q$
and the $z$-component of the total spin $\hat S_z^t$. Still, the Hamiltonian $H_M^0$ can be diagonalized exactly and
the states constructed in an analytical fashion in terms of the states introduced previously for $J=0$,
by using the Clebsch-Gordan construction \cite{Timm2006}. Now the new states $|Q,m\rangle$ shall be classified by the
molecular charge $Q$, and by the  $z$-component of the total spin, $m$.

For $m\in [-S+1/2,S-1/2]$ the single-particle states ($Q=1$) are given by:
\begin{equation}\label{states}
|1,m\rangle^{\pm}= C_{m\downarrow}^{\pm}|\downarrow,m+1/2 \rangle+C_{m\uparrow}^{\pm}|\uparrow,m-1/2 \rangle,
\end{equation}
and their associated eigenvalues ${\cal E}_{1,m}^{\pm}$ read as:
\begin{equation}\label{Eigen}
{\cal E}_{1,m}^{\pm}=\epsilon-g\mu_B Bm+\frac{J}{4}-D\left ( m^2+\frac{1}{4}\right )\pm\Delta {\cal E}(m),
\end{equation}
where $\Delta {\cal E}(m)=[D(D-J)m^2+(J/4)^2(2S+1)^2]^{1/2}$.
The coefficients $C_{m\sigma}^{\pm}$ in Eq~(\ref{states}) are simply the Clebsch-Gordan coefficients.
The states $|0,S_z\rangle$ are not affected by the exchange coupling and one has $|0,m\rangle=|0,S_z\rangle$.
The corresponding eigenvalue is simply ${\cal E}_{0,m}=-DS_z^2-g\mu_B BS_z$. The remaining $Q=1$ states are
$|1,-S-1/2\rangle$ and $|1,S+1/2\rangle$. For a vanishing magnetic field, $B=0$, the states associated to $\pm m$ are degenerate and one has
\begin{eqnarray}\label{degeneracy}
{\cal E}_{1,m}^{\pm}(B=0)&=&{\cal E}_{1,-m}^{\pm}(B=0),\\
{\cal E}_{0,m}(B=0)&=&{\cal E}_{0,-m}(B=0).
\end{eqnarray}

So far we have discussed how to construct the states and to compute the energy spectrum for $H_M^0$.
In the rest of this paragraph we shall address the role of the transverse anisotropy term.
The transverse anisotropy term $\sim ({\hat S}_x^2-{\hat S}_y^2) \sim ({\hat S}_+^2+{\hat S}_-^2)$
does not commute with $\hat S_z^t$ and induces transitions \cite{MolecularNanomagnets} between the states of $H_M^0$ with the selection rule
$|m-m'|=2$.

As the molecular charge is a good quantum number even in the presence of the transverse anisotropy,
 the eigenstates of the total molecular Hamiltonian $H_M$ can be classified by the molecular charge $Q$ only. We shall label
them $|\varphi_{Q,\nu}\rangle$, where $Q=\{0,1\}$  (states with molecular charge $Q=2$ are disregarded) while $\nu$ is an internal
label that indexes the states within a multiplet. In the presence of the transverse anisotropy $E$, the
`empty' molecular states (EMS) can be written as:
\begin{equation}\label{EMS}
|\varphi_{0,\nu}\rangle=\sum_m c_{\nu,m}|0,m\rangle, \quad \nu=1,..,2S+1,
\end{equation}
with $m$ running over all allowed values in the range $[-S,S]$.
For half integer $S$ and a small magnetic field, the transverse anisotropy plays a minor role in the mixing
of the states $|\varphi_{0,\nu}\rangle$, as the transition amplitudes between the empty molecular states
$\{|0,m\rangle\}$ are negligible (see also the discussion following Fig.~\ref{fig2}.)

In contrast, the transverse anisotropy couples the degenerate, one-particle states  ($Q=1$) with opposite $m$'s.
The strongest mixing is expected for the pairs $|1,1\rangle^{\pm}$ and $|1,-1\rangle^{\pm}$  as the off-diagonal
matrix element $^{\pm}\langle1,1|H_M|1,-1\rangle^{\pm}$ is linear in $E$. Higher order mixing effects become important
as the ratio $E/D$ increases and one can generally write:
\begin{eqnarray}\label{nu-mix}
|\varphi_{1,\nu}\rangle&=&\sum_{m}\sum_{s}c_{\nu,m}^s|1,m\rangle^s, \quad \nu=1,..,2(2S+1).
\end{eqnarray}
The eigenvalues $E_{Q,\nu}$ of $H_M$ and the coefficients in Eqs.(\ref{EMS}) and (\ref{nu-mix}) can be
 found only by numerical diagonalization. More details on the spectral properties and state mixing  will be given in Section 3.

Finally, we write down the matrix elements of the tunneling Hamiltonian $H_T$ with respect to the eigenstates of $H_M^0$
and derive the selection rules for molecular transitions due to electronic back-and-forth processes namely
($\lambda_{\uparrow}=1$, $\lambda_{\downarrow}=-1$):
\begin{equation}\label{Srules}
^{\pm}\langle 1,m|a^{\dagger}_{\sigma}|0,m'\rangle = C_{m\sigma}^{\pm}\delta_{m,m'+\lambda_{\sigma}/2}.
\end{equation}
This equation describes the tunneling of one electron with spin $\sigma$ on the SMM orbital,
when the number of electrons in the molecule increases by one while the {\it total} magnetic quantum number $m$
changes by $\pm 1/2$.

\subsection{Generalized master equation approach}

The GME approach which we use
to investigate the time-dependent transport relies on the partitioning approach \cite{Caroli}.
More precisely, transient currents develop in the source and drain electrodes as they are contacted to the molecule at
some initial instant. The leads are viewed as non-interacting particle reservoirs with chemical potentials $\mu_{\alpha}$ ($\alpha=L,R$),
and at equilibrium described by the Hamiltonian (\ref{eq:H_leads}).
This setting is suitable for perturbative calculations with respect to the lead-molecule couplings $\Gamma^{\alpha}_{\pm}$, and allows us
to compute transient currents in the presence of time-dependent modulation of the contacts, as in the turnstile regime.

 The GME method essentially provides the SMM reduced density operator (RDO) $\rho$ defined as the partial trace over
the leads' degrees of freedom $\rho(t)=\rm {Tr}_{{\cal F}_{\rm el}}\{W(t)\}$. Here $W(t)$ is the density operator of the whole
structure which solves the Liouville - von Neumann equation $i\hbar{\dot W}(t)=[H(t),W(t)]$, and the trace is over ${\cal F}_{{\rm el}}$ - the Fock space
of the non-interacting electronic reservoirs. In the sequential tunneling regime considered here the master equation takes a
rather compact form (for a full derivation see Ref. \cite{Moldo1}):
\begin{eqnarray}\label{GME}
\frac{d\rho(t)}{dt}&=&-\frac{i}{\hbar}[H_M,\rho(t)]-\frac{1}{\hbar^2}{\rm Tr}_{{\cal F}_{{\rm el}}}\left\lbrace
{\cal D}_t[\rho]\right \rbrace,\\\label{D}
{\cal D}_t[\rho]&=&[H_T(t),\int_0^t\,dsU_{t-s}[H_T(s),\rho(s)\rho_{{\rm el}}]U_{t-s}^*],
\end{eqnarray}
where we introduced the ``free'' evolution operator of the disconnected system $U_{t}=e^{-i(H_M+H_L+H_R)t/\hbar}$
and the equilibrium distribution of the leads $\rho_{{\rm el}}$ \cite{Fermi}. The dissipative operator ${\cal D}_t$ collects all
sequential tunneling processes from the switching instant $t_0=0$ to the current time $t$. We solve numerically Eq.~(\ref{GME})
with respect to the fully interacting states of $H_M$ and obtain the populations associated
to a given state $|\varphi_{Q,\nu}\rangle$ as
\begin{equation}\label{population}
P_{Q,\nu}(t)=\langle \varphi_{Q,\nu}|\rho(t)|\varphi_{Q,\nu}\rangle .
\end{equation}
Once $\rho(t)$ is known one can calculate average values of molecular observables by performing a trace over
the Fock space ${\cal F}_M$ of the molecule. For instance, the total charge accumulated on the orbital involved in transport
is given by
\begin{equation}
 Q(t)=e{\rm Tr}_{{\cal F}_M}\lbrace \rho(t) {\hat N}\rbrace,
\end{equation}
where the total electronic occupation ${\hat N}={\hat n}_{\uparrow}+{\hat n}_{\downarrow}$ and $e$ is the electron charge.
The continuity equation then becomes \cite{PRB-I}
\begin{equation}\label{CEQ}
J_{L}(t)-J_R(t)=\sum_{Q}\sum_{\nu_Q}\langle \varphi_{Q,\nu_Q}|\dot\rho(t)|\varphi_{Q,\nu_Q}\rangle,
\end{equation}
where $\nu_{Q}$ is the set of states with charge $Q$ .
By inserting the tunneling Hamiltonian $H_T$ into the double commutator ${\cal D}_t$ given in Eq.(\ref{D})
one identifies $J_L$ and $J_R$ from the RHS of Eq.(\ref{CEQ}). It is straightforward to show that the `empty' molecular states
$|\varphi_{0,\nu}\rangle $ do not contribute to the currents.
Similarly, one can calculate the total spin $\langle S^t_z\rangle={\rm Tr}_{{\cal F}_M}\lbrace \rho(t) {\hat S}_z^t\rbrace $ as well as the spin currents.
In this work the relaxation of the excited molecular states via phonon emission is not considered.
This is a good approximation as long as the timescale on which the quantum turnstile operates is much
smaller than the relaxation time which is of order of $10^{-6}$ s (see e.g Refs. \cite{Ardavan,Bahr}).
 In fact previous studies \cite{MB2008} reported that the  current-induced magnetic switching is stable against
intrinsic spin-relaxation processes.

\section{Results and discussion}\label{sec:Results}

\subsection{The transport configuration and tunneling processes}\label{sec:QT}

\begin{figure}[tbhp!]
\begin{center}$
\begin{array}{cc}
\includegraphics[width=6.55cm]{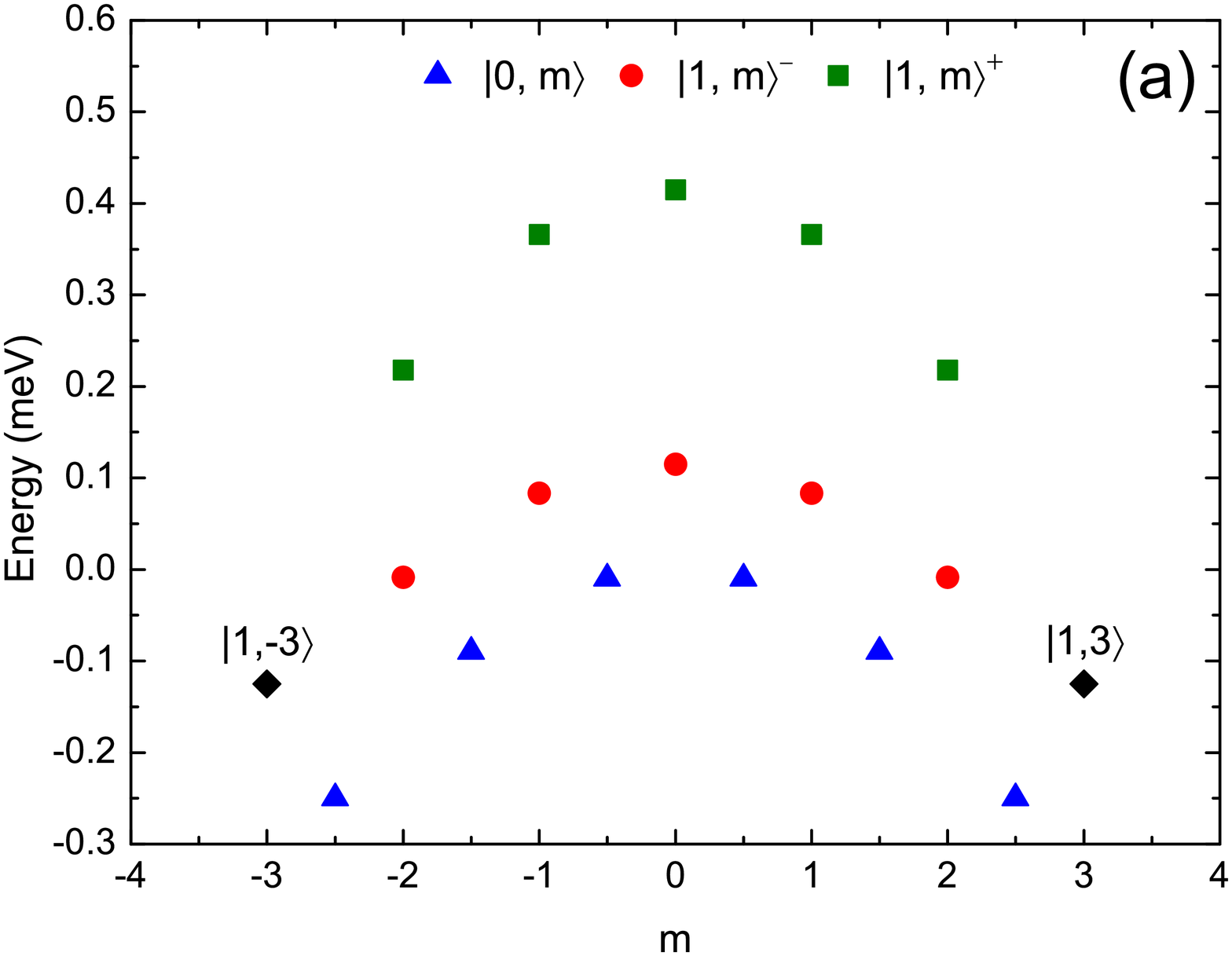} &
\includegraphics[width=7.5cm]{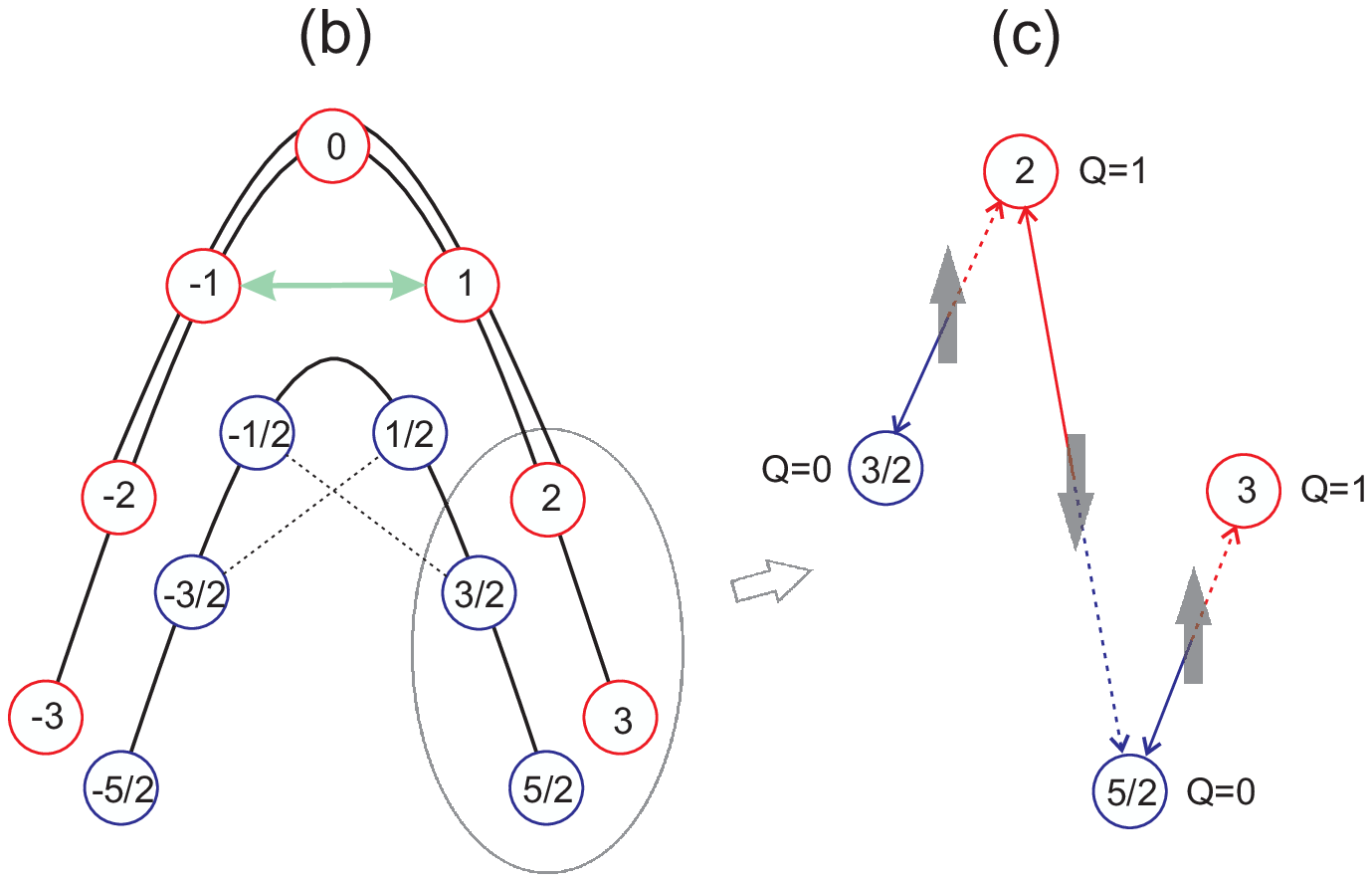}
\end{array}$
\end{center}
\caption{(Color online) (a) The energy levels for a SMM of spin $S=5/2$ as function of the total magnetic quantum number $m$.
The two-particle sector is not included as the corresponding states are outside the selected bias window.
Other parameters are: $\epsilon=0.25$meV, $J=0.1$meV, $U=1$meV, $D=0.04$meV, $g\mu_B B=0$ and $E=0$.
(b) The transverse anisotropy $E$ induces mixing of degenerate or nearly degenerate states
with $m=1$ and $m=-1$ (indicated by the double arrow). There is no significant quantum tunneling of magnetization between
the states $|0,\mp1/2\rangle$ and $|0,\pm 3/2\rangle$ (indicated by the dashed lines). The numbers in the circles are the total magnetic 
quantum numbers $m$. (c) Tunneling processes connecting the lowest four states of the SMM via back-and-forth tunneling with the leads.
These states are also involved in the first turnstile cycle - see the discussion in the text.  }
\label{fig2}
\end{figure}

The numerical simulations were performed for molecules with spin $S=5/2$ but our conclusions remain valid for larger
half-integer values of $S$.
For reasons that will become clear below in this work we restrict ourselves to SMMs with small transverse anisotropy,
that is $E\ll D\ll J$
(in fact we allow a maximum ratio $E/D=1/25$ at fixed easy-axis anisotropy constant $D$).
We shall investigate two spin configurations for the leads. In the so called antiparallel
(AP) configuration the left lead carries only spin-down electrons and the right lead is spin-up polarized.
For the second configuration the left lead is non-magnetic (i.e. $P^L=0$) and the right lead remains ferromagnetic.
We label this configuration as normal-ferromagnetic (NF).

 The chemical potential of the leads $\mu_{L,R}$ are set such that only the states $|\varphi_{0,\nu}\rangle$ and $|\varphi_{1,\nu}\rangle$ contribute to the tunneling processes,
while double occupied states, $|\varphi_{2,\nu}\rangle$,
have a much higher energies, $ E_{2,\nu} > \mu_{L}$, and do not contribute to transport.

Let us first discuss the energy spectrum and the relevant lead-molecule tunneling
processes {\it in the absence of transverse anisotropy}.
In the following discussion we shall use the basis $\{|Q,m\rangle\}$ of $H_M^0 = H_{M} (E=0)$.
As we are interested in the pumping mechanism at finite transverse
 anisotropy $E$, we shall switch later to the basis $|\varphi_{Q,\nu} \rangle$ of the full $H_M$. In that situation, 
with certain modifications, a similar turnstile scenario holds.
Figure\,2(a) shows the energy levels of $H_{M}^0$ for a given set of parameters, and in the absence of the
external magnetic field. Figure \,2(b) schematically shows the charge $Q=1$ (integer $m$'s) and $Q=0$ (half integer $m$'s) branches of the spectrum; 
in view of further discussion the double arrow and
the dotted lines mark some of the quantum tunneling of magnetization (QTM) processes induced by a nonvanishing $E$.
The states connected by the double arrow are strongly mixed by $E$, while the ones connected by the
dashed lines are only weakly coupled.
As charge $Q$ is conserved, it implies that  direct transitions
between $Q=0$ and $Q=1$ branches are forbidden by symmetry. This is only possible through
processes involving states in the leads that do not conserve the charge.
For example in Fig.~\ref{fig2}(c), we show such processes (blue and red arrows). The same figure also shows
how the SMM evolves from an initial `empty' molecular state $|0,5/2\rangle$ to the to the next EMS $|0,3/2\rangle$ via tunneling processes.

We shall call the transitions $m\to m-1/2$ `forward' processes (they follow the full line arrows in Fig.~\ref{fig2}(c))
as they contribute to the magnetic switching $m=5/2\to m=-5/2$. On the other hand
the transitions $m\to m+1/2$ compete for the total spin reversal and can be regarded as `backward' processes
(they follow the dashed lines in Fig.~\ref{fig2}(c)).

Furthermore, we distinguish two types of `forward' transitions: (i) `absorption' of spin-down
electrons from the leads, i.e. the charging of the molecular orbital along the transitions
$|0,m\rangle\to|1,m-1/2\rangle^{\pm}$ (full red arrow in Fig.~\ref{fig2}(c)) and (ii) tunneling of spin-up electrons
from the molecular orbital, i.e. a depletion process associated to the transitions $|1,m-1/2\rangle^{\pm}\to|0,m-1\rangle$
(full blue arrow).
Similarly one defines charging and discharging `backward' processes (associated with the dashed lines in Fig.~\ref{fig2}(c)).
We find this analysis useful as it provides hints for a write-and-read scheme of states with well defined molecular
spin $|0,m\rangle$ when operating the SMM in the turnstile regime. Such a protocol will be discussed in the next subsection.

\subsection{The turnstile protocol}\label{sec:turnstile}

 A turnstile pumping cycle entails two steps: (i) the charging of the molecular orbital from the left lead while
the drain contact is closed ($\chi_L(t)\neq 0$, $\chi_R(t)=0$) and (ii) discharging/depletion through the drain
lead ($\chi_L(t)= 0$, $\chi_R(t)\neq0$). We simulate the turnstile operation by appropriately tailoring the switching
 functions $\chi_{L}$ and $\chi_{R}$ in the tunneling Hamiltonian $H_T$.

The main idea behind the proposed operation is the following: use the charging sequences to prepare
intermediate one-electron states via `forward' tunneling from the left lead and write `empty' molecular
states $|0,m\rangle$ along `forward' depletions to the right lead.
To be more precise, let us discuss a single turnstile cycle at $E=0$
in the NF configuration, starting from the initial state $|0,5/2\rangle$.
The associated transitions are depicted in Fig.~\ref{fig2}(c). By opening the source (left) contact, 
the states $|1,2\rangle^{\pm}$ become populated by absorbing one spin -- $\downarrow$ electron 
(forward tunneling). In turn, if a spin -- $\uparrow$ electron
is absorbed then the rightmost state $|1,3\rangle$ becomes activated (backward tunneling).
Then, the left contact closes and the drain (right) electrode comes into play.
Now, the orbital is depleted through forward tunneling $|1,3\rangle\to|0,5/2\rangle$ and 
$|1,2\rangle^{\pm}\to |0,3/2\rangle$, as spin -- $\uparrow$ electron tunnels out into the right lead.
An accurate operation would lead to the preparation of a {\it single} EMS or to an equally weighted mixture of EMS,
but this scenario is not expected to work if the transverse anisotropy induces strong mixing of states $|0,m\rangle$.

 In view of this analysis, let us now discuss how this picture gets modified in the
 presence of the transverse anisotropy. We start by describing the construction of
the empty molecular states  $\{|\varphi_{0,\nu}\rangle\}$.
We find that if $g\mu_B B\ll D$ the mixing of `empty' molecular states $|0,m\rangle$ is negligible since the
QTM between the states $|0,\mp1/2\rangle$ and $|0,\pm 3/2\rangle$ is very weak (see the dotted lines in Fig.\,2(b)).
In this case we find a one to one correspondence between $\nu\leftrightarrow m$ as
for any $\nu$ in Eq.(\ref{EMS}) one can find a single $m$ such that $|\varphi_{0,\nu}\rangle\approx |0,m\rangle$.

This simple correspondence fails as the magnetic field approaches resonant value $g\mu_B B_{{\rm res}}=-D(S_z+S_z')$,
and the Landau-Zenner tunneling processes between ${\cal E}_{0,m}$ and ${\cal E}_{0,m'=m\pm 2}$ become
important and lead to strong mixing of the states. Such a resonant regime will not be discussed in
the present work.

We now turn to $Q=1$ states. For $B=0$, the states $|1,m\rangle^{\pm}$ and
$|1,-m\rangle^{\pm}$ in Eq.(\ref{nu-mix}) are mixed by the transverse anisotropy term as
${\cal E}_{1,m}^{\pm}$ and ${\cal E}_{1,-m}^{\pm}$ are degenerate (see Eq.(\ref{degeneracy})).
The mixing is indicated by the double arrow in Fig.~\ref{fig2}(b).
 However, even a small magnetic field lifts this degeneracy and one finds a rather small mixing of the states
$|1,m\rangle^{\pm}$ and $|1,-m\rangle^{\pm}$ for $E\ne 0$. Once again, for
each $|\varphi_{1,\nu}\rangle$ there is  a single state $|1,m'\rangle^s$ of $H_M^0$ whose weight
$|c_{\nu,m'}^s|^2$ in Eq.(\ref{nu-mix}) is by far the largest one.
Under these conditions the one-to-one correspondence between the index $\nu$ and the quantum number $m$ is preserved
for all states and allows us to index them as:
\begin{equation}\label{short}
|\varphi_{1,\nu}\rangle\to |\varphi^s_{1,m}\rangle \approx |1,m\rangle^s,  \quad
|\varphi_{0,\nu}\rangle \to |\varphi_{0,m}\rangle \approx  |0,m\rangle.
\end{equation}
We shall close this section by noticing that although this representation works, $\nu$ will always be read as
an index, and not as a quantum number.
Consequently, in the ${|\varphi^s_{1,m}\rangle}$ basis the populations of the states
will be denoted by $P_{|\varphi^s_{1,m}\rangle}$ and $P_{|\varphi_{0,m}\rangle}$.

\subsection{Writing and reading the excited molecular states}\label{sec:WR}
\begin{figure}[tbhp!]
\includegraphics[angle=0,width=0.85\textwidth]{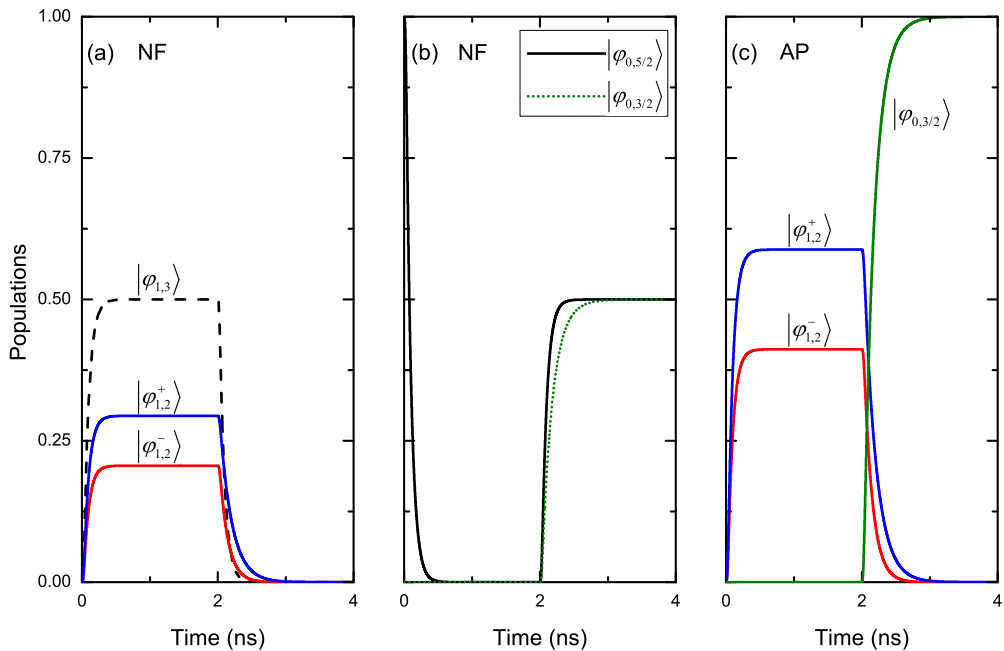}
\centering
\caption{(Color online) The evolution of the relevant populations along the first turnstile cycle
for NF an AP configurations. The charging sequence corresponds to $t\in [0,2]$ ns and the depletion sequence
 to $t\in [2,4]$ ns. (a) and (b) - normal-ferromagnetic (NF) configuration, (c) - antiparallel (AP) configuration;
the system ends up in a single excited state $|\varphi_{0,3/2}\rangle$ (see also the discussion in the text).
Other parameters: $\epsilon=0.25$meV, $J=0.1$meV, $U=1$meV, $\mu_L=1$meV, $\mu_R=-1$meV, $D=0.04$meV, $g\mu_BB=$0.005meV
and $E/D=1/250$. }
\label{fig3}
\end{figure}
\begin{figure}[tbhp!]
\includegraphics[angle=0,width=0.55\textwidth]{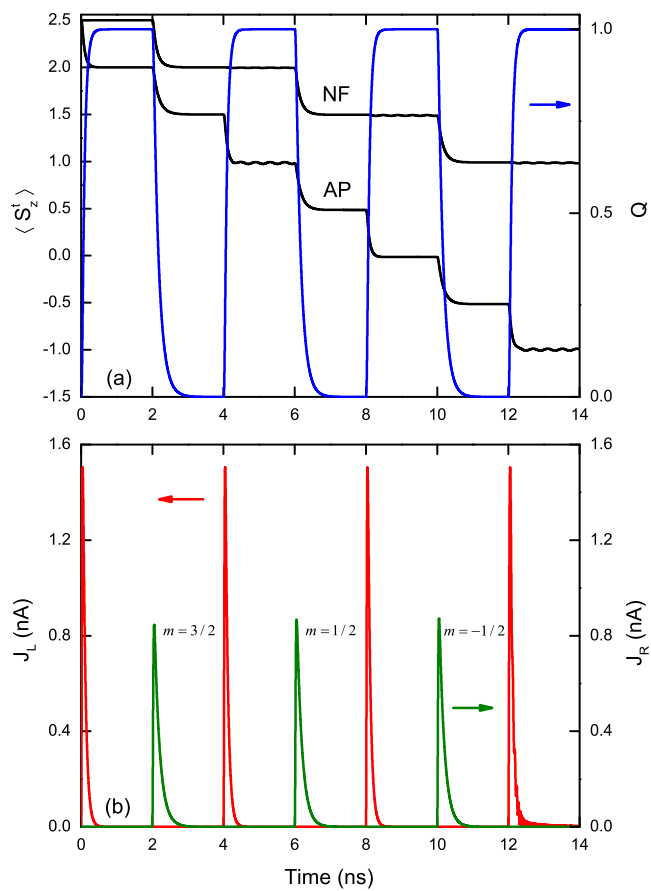}
\centering
\caption{(Color online) (a) The average total molecular spin $\langle S_z^t\rangle$ in the anti-parallel (AP)
and normal-ferromagnetic (NF) configurations - black lines.
The step-like structure is discussed in the text. The total charge $Q$ accumulated on the molecular orbital - blue line.
(b) The transient currents $J_{L,R}$ in the AP configuration. The half-integer steps of $\langle S_z^t\rangle$ suggest that in the
corresponding time range the SMM state is simply $|0,m\rangle$ - see the discussion in the text. The pumping period is 2ns.
Other parameters: $\mu_L=1$\,meV, $\mu_R=-1$\,meV, $\epsilon=0.25$\,meV, $J=0.1$\,meV, $U=1$\,meV
and $\tau=0.5$\,meV, $D=0.04$\,meV, $E/D=1/250$, $g\mu_B B=0.005$\,meV, $V^L=V^R=0.045$\,meV, $k_BT=0.001$\,meV.}
\label{fig4}
\end{figure}

We performed transport calculations starting from the initial state  $|\varphi_{0,5/2}\rangle$,
so the density matrix  describing the system at $t=0$ is
$\rho(t=0)=|\varphi_{0,5/2}\rangle\langle\varphi_{0,5/2}|$.
As stated previously, we shall present results for small values of the ratio $E/D\sim 10^{-2}$.
The evolution of the relevant populations along a {\it single} turnstile cycle in the normal-ferromagnetic (NF)
configuration for $E/D=1/250$ is presented in Fig.~\ref{fig3}(a) and (b). The tunneling processes are similar
to the ones discussed along Fig.~\ref{fig2}(c) when $E=0$. The state $|\varphi_{1,3}\rangle$
 is half filled through spin-up `backwards' tunneling whereas $P_{|\varphi^+_{1,2}\rangle}+P_{|\varphi^-_{1,2}\rangle}=1/2$.
The small imbalance population of the states $|\varphi^{\pm}_{1,2}\rangle$ is due to the finite $J$, while
$P_{|\varphi^+_{1,2}\rangle}=P_{|\varphi^-_{1,2}\rangle}$  at $J=0$.
 Along the charging transition towards the $Q=1$ sector, the population
$P_{|0,5/2\rangle}$ drops quickly to zero. The depletion cycle $t\in [2,4]$ ns
simultaneously activates the states $|\varphi_{0,5/2}\rangle$ and $|\varphi_{0,3/2}\rangle$,
the stationary regime being described by the RDO
$\rho=(|\varphi_{0,5/2}\rangle\langle \varphi_{0,5/2}|+|\varphi_{0,3/2}\rangle\langle \varphi_{0,3/2}|)/2$.
Therefore one can use the NF configuration to prepare an equally weighted mixture of states. 
Along the first depletion sequence, in the NF configuration,
$\langle S_z^t\rangle=2$ (see Fig.~\ref{fig4}(a)).

In the AP configuration, the first turnstile cycle drives the SMM out of the ground state
$|\varphi_{0,5/2}\rangle$ directly into the first excited `empty' molecular state
$|\varphi_{0,3/2}\rangle$ (see Fig.~\ref{fig3}(c)), with no further mixing as in NF configuration.
The population of this excited state attains its maximum value within the depletion cycle.
In the AP configuration the two EMSs can be viewed as binary digits
(i.e. $|\varphi_{0,5/2}\rangle\to |0\rangle$ and $|\varphi_{0,3/2}\rangle\to |1\rangle$)
that are switched along the turnstile protocol.

The writing of a single EMS depends crucially on the spin polarization of the leads.
 Fig.~\ref{fig3}(c) indicates  that the `backward' processes (both charging and relaxation)
are forbidden, as the state $|\varphi_{1,3}\rangle$ is not available along the charging sequence
because the left lead provides no spin-up electrons. For the same reason there are no transitions
from $|\varphi^{\pm}_{1,2}\rangle$ back to $|\varphi_{0,5/2}\rangle$ on the discharging sequence.
In fact, on the charging cycle the system occupies only two states with imbalanced populations
$P_{|\varphi^+_{1,2}\rangle}> P_{|\varphi^-_{1,2}\rangle}$ and
 which eventually deplete in favor of $|\varphi_{0,3/2}\rangle$. It is important to observe
that on the depletion cycle the average total spin $\langle S_z^t\rangle=3/2$ and coincides with the molecular spin of the
$|\varphi_{0,3/2}\rangle$ EMS (see Fig.~\ref{fig4}(a)).

The reverse magnetic switching can be also implemented by simply reversing the bias
($\mu_L\leftrightarrow\mu_R$) while keeping both contacts closed and then repeating the turnstile
operation with the new initial state $|\varphi_{0,3/2}\rangle$. Then the system returns to $|\varphi_{0,5/2}\rangle$.
This is a {\it classical} NOT operation, as the system evolves from $|\varphi_{0,5/2}\rangle$ to $|\varphi_{0,3/2}\rangle$ and then back to $|\varphi_{0,5/2}\rangle$ without passing through a superposition of these states.

Let us emphasize that the preparation of a pure excited molecular state $|\varphi_{0,m}\rangle$ cannot be achieved in
the standard transport regime. In that case the charge flows simultaneously to and from the
 leads and one cannot completely deplete the molecule and therefore $\langle \varphi_{0,m}|\rho(t)|\varphi_{0,m}\rangle<1$. It should be mentioned that for larger $S$ the time needed to achieve the
full magnetic switching $m=S\to -S$ also increases as
 the system must visit all the intermediate states~\cite{Timm2006}. This fact suggests that the pair of consecutive states $(|\varphi_{0,S}\rangle,|\varphi_{0,S-1}\rangle)$
might be more appropriate for faster manipulation of magnetic qubits.

Given these results one can ask about the time evolution of the total spin under repeated pumping
cycles and on the possibility to read the states prepared along the turnstile operation by measuring
 currents.
Fig.~\ref{fig4} summarizes our main results
on transient currents and spin evolution along few turnstile cycles for the AP and NF configurations.
The time-dependent occupation of the molecular orbital (the blue line in Fig.\,4(a)) has a
typical charging/relaxation pattern, with quick orbital filling and slightly slower depletion.
This can be seen by comparing the abrupt increase (in less than 1/2 ns) of the population at the beginning of the 
charging cycles (e.g. $t=4,8,12$ns) to the smooth tail of discharging  which extends over 1ns
(e.g the time range $[6,7]$ns).

The total spin average $\langle S_z^t\rangle$ presented in  Fig.\,4(a) displays a
step-like structure in both configurations. The steps scan both integer and half-integer values of $\langle S_z^t\rangle$,
the last step for the AP configuration corresponding to $\langle S_z^t\rangle=3$ being reached after $t\simeq18$\,ns (not shown).
 In the AP configuration the onset of half-integer steps corresponds
to the depletion of the molecular orbital ($Q=1\to Q=0$), whereas the transition between half-integer to integer steps is associated to the charging process ($Q=0\to Q=1$).

The NF configuration presents different features: there are fewer but longer steps of $\langle S_z^t\rangle$, but no clear
correspondence can be made between these steps and the behavior of the
total charge $Q$. One notices that in this case the integer steps extend on some charging sequences and that half-integer
values are encountered even if the orbital is empty.
Fig.~\ref{fig4}(b) displays the expected series of spikes for the  transient currents $J_{L,R}$ in the AP configuration.
The period of the pumping cycles must be chosen appropriately in order to ensure full charging and discharging of
the molecular orbital (we find the minimal period to be $\sim$ 1 ns).

We have found a similar behavior (not shown here) for the transient currents in the NF configuration.
The input current $J_L$ vanishes when the orbital is fully occupied ($Q=1$),
whereas on the discharging sequence $J_R$ drops to zero as the orbital depletes.
The amplitudes of $J_L$ and $J_R$ are different
because the charging process is faster than the depletion (see Fig.~\ref{fig4}(a)).
By inspecting Figs.~\ref{fig4}(a) and (b) one observes that
in the AP configuration we have a one to one correspondence between the average $\langle S_z^t\rangle$
and the peak-to-peak sequence in the
transient currents: $\langle S_z^t\rangle$  acquires half-integer values only between a depletion 
peak and the next charging peak (e.g. for $t\in [2, 4]$ns $\langle S_z^t\rangle=3/2$), while between 
a charging peak and the next depletion peak the average spin is an integer.
This means that the AP configuration can be used to
record experimentally the initialization of a given `empty' molecular state $|\varphi_{0,m}\rangle$.

To this end it is sufficient to know the initial state of the molecule and to carefully `count' the
transient peaks of $J_L$ and $J_R$. We need to keep in mind though that Fig.~\ref{fig4}(a) shows the {\it average}
value of the total spin, which does not guarantee that along half-integer steps of $\langle S_z^t\rangle$
the system is in a pure state characterized by the RDO $\rho\sim |0,m\rangle\langle 0,m|$, especially
for larger values of the transverse anisotropy when one expects stronger mixing of states.

\subsection{Transverse ansiotropy effects}\label{sec:E-effects}

\begin{figure}[tbhp!]
\includegraphics[angle=0,width=0.85\textwidth]{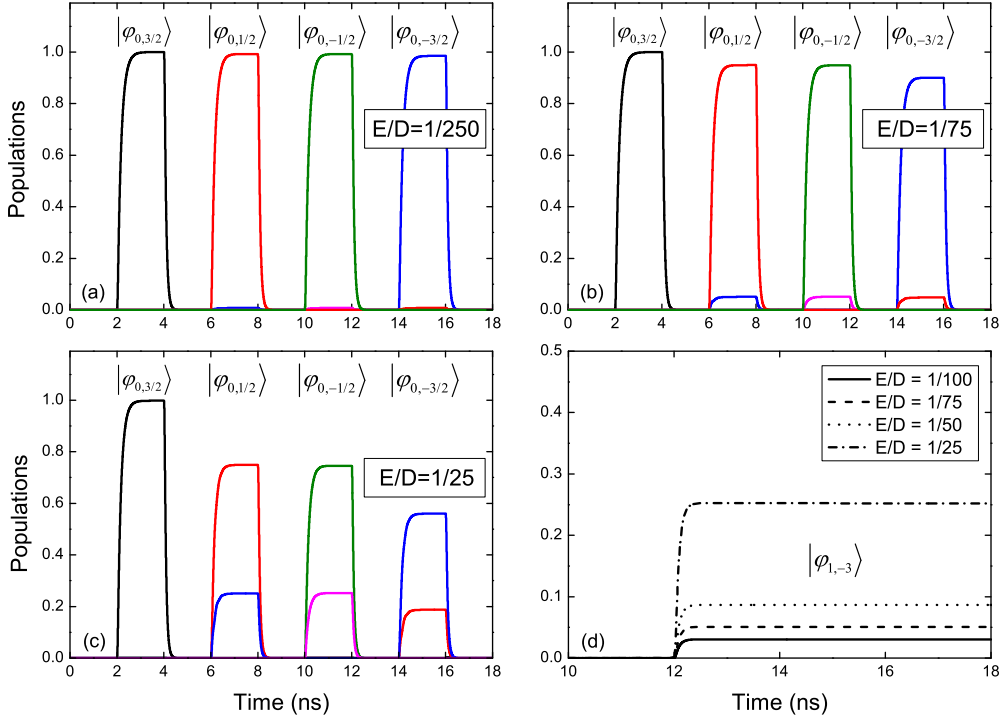}
\centering
\caption{ (Color online) The populations $P_{|\varphi_{0,m}\rangle}$ of the empty molecular states
in the antiparallel configuration. Black line - $P_{|\varphi_{0,3/2}\rangle}$, red line - $P_{|\varphi_{0,1/2}\rangle}$,
green line - $P_{|\varphi_{0,-1/2}\rangle}$, blue line - $P_{|\varphi_{0,-3/2}\rangle}$,
magenta line - $P_{|\varphi_{0,-5/2}\rangle}$.
(a) $E/D=1/250$. (b)  $E/D=1/75$. (c)  $E/D=1/25$. On each turnstile cycle we indicate the dominant EMS.
(d) $P_{|\varphi_{1,-3}\rangle}$ for different values of the ratio $E/D$. The other parameters are the same with those in Fig.\,3.}
\label{fig5}
\end{figure}

To further investigate the role of the anisotropy,
we calculated the populations $P_{|\varphi_{0,m}\rangle}$ of several `empty' molecular states $|\varphi_{0,m}\rangle$
for different values of the ratio $E/D$, at fixed magnetic field. Fig.\,5(a) confirms that at $E/D=1/250$
the $k$-th depletion cycle is well described by a single state $|\varphi_{0,m=5/2-k}\rangle$.
This proves the stepwise all-electrical writing of EMS (i.e. point (i) in the Introduction).

By increasing the transverse anisotropy constant such that $E/D=1/75$ we notice in Fig.~\ref{fig5}(b) the 
emergence of a 2nd EMS on the depletion cycles. Nevertheless, the population of the dominant `empty' molecular 
state exceeds 0.9 so we can still associate a well defined molecular state to {\it each} of the depletion cycle.
 This no longer holds for $E/D=1/25$. Fig.~\ref{fig5}(c) reveals that the weight of the state 
$|\varphi_{0,-3/2}\rangle$ and $|\varphi_{0,-5/2}\rangle$ on the 2nd and 3rd depletion cycle increases up to 0.25,
reducing the efficiency of the quantum turnstile protocol. 
Moreover, one can easily see that along the 4th cycle (i.e $t\in [14,16]$ns) 
$P_{|\varphi_{0,-3/2}\rangle}+P_{|\varphi_{0,1/2}\rangle}<1$ which suggests that other states have to be populated.
We have found that the  state $|\varphi_{1,-3}\rangle$, corresponding to $Q=1$ gets populated after the third cycle 
due to the forward transition $|\varphi_{0,-5/2}\rangle\to|\varphi_{1,-3}\rangle$ via spin -- $\downarrow$ tunneling
into the SMM.  

\begin{figure}[tbhp!]
\begin{center}$
\begin{array}{cc}
\includegraphics[width=7.55cm]{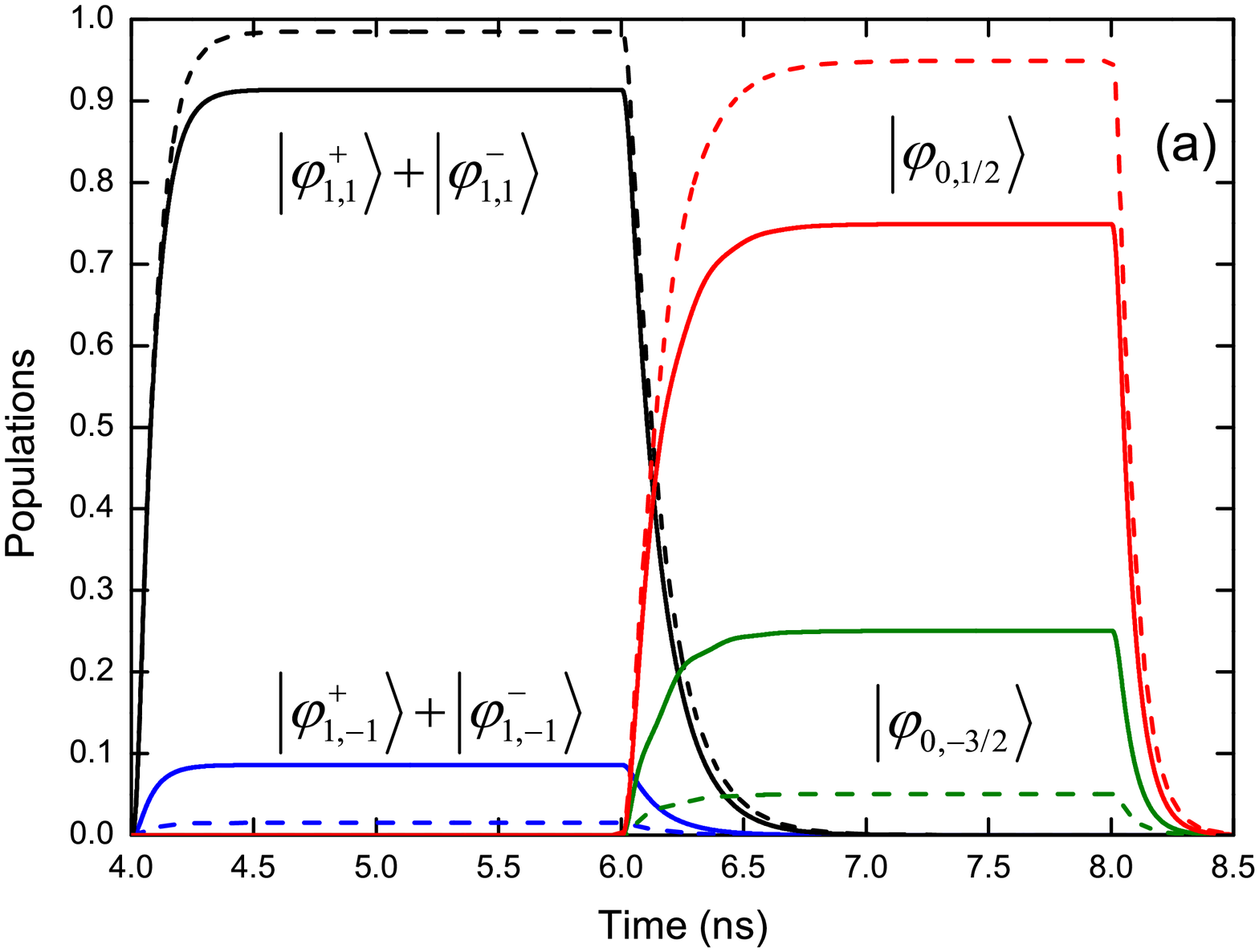} &
\includegraphics[width=6.45cm]{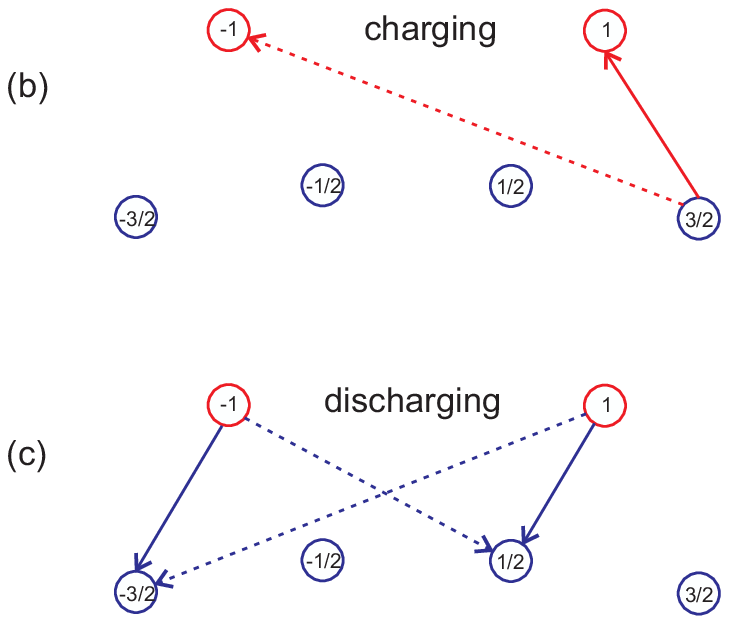}
\end{array}$
\end{center}
\caption{ (Color online) (a)  The populations of active states along the second turnstile cycle for two values of the
anisotropy constant. The dashed lines correspond to $E/D=1/75$ and the solid lines to an increased value $E/D=1/25$.
More discussion is given in the text. Other parameters are as in Fig.\,3. (b) and (c): Schematic representation
of the tunneling processes between the states $|\varphi^{\pm}_{1,\pm 1}\rangle$ and EMSs along the
charging and discharging sequences. The numbers denote the dominant total magnetic moment $m$ of the
fully interacting one-particle states. }
\label{fig6}
\end{figure}

In Fig.~\ref{fig5}(d) we show that the population of this state becomes relevant with increasing the 
anisotropy, i.e. from a population of 0.03 at $E/D=1/100$ to 0.25 at $E/D=1/25$.

In order to explain the coexistence of two EMSs on the same depletion sequence when the transverse anisotropy
increases we have to analyze the  QTM between nearly degenerate  $Q=1$ states.
By looking at the off-diagonal matrix element $^{\pm}\langle 1,1|({\hat S}_+^2+{\hat S}_-^2)|1,-1\rangle^{\pm}$ we
infer that by increasing $E$ the hybridization of $|1,1\rangle^{\pm}$ and $|1,-1\rangle^{\pm}$
in the fully interacting states increases as well. We find that the weight of the `minority' state
$|1,-1\rangle^{\pm}$ in $|\varphi^{\pm}_{1,1}\rangle$ increases from $\sim 10^{-2}$ for $E/D=1/250$ to $\sim 10^{-1}$ for $E/D=1/25$.
The mixing of the states $|1,2\rangle^{\pm}$ and $|1,-2\rangle^{\pm}$ arises to the second order in $E$ and is still negligible.
As a consequence the accuracy of the first turnstile cycle is preserved even for $E/D$ as large as $E/D\simeq 1/25$ and
that $P_{|\varphi_{0,3/2}\rangle}\approx 1$. This is confirmed by the results presented in
 Fig.~\ref{fig5}(a)-(c). In order to recover `clean' EMSs on each depletion cycle for larger values 
of $E/D$ one could slightly increase the magnetic field. The latter lifts even more the degeneracy of the states 
$|1,1\rangle^{\pm}$ and $|1,-1\rangle^{\pm}$ and reduces therefore their mixing in the presence of $E/D$. 

Fig.~\ref{fig6}(a) presents the relevant populations of EMSs
and $Q=1$ states
on the second turnstile cycle ($t\in [4,8]$ns) for two values of the ratio $E/D$. For simplicity we plot the total
population of states corresponding to the same dominant value of spin $m$. 
Figs.~\ref{fig6}(b) and (c) indicate schematically
the relevant tunneling processes between states of $H_M$ along the charging and discharging sequences.
The states $|\varphi^{\pm}_{1,\pm 1}\rangle$ are simultaneously filled along the charging sequence.
On the other hand, the states $|\varphi^{\pm}_{1,-1}\rangle$ are less responsive to charging
(this processes correspond to the dashed line in Fig.~\ref{fig6}(b)) because spin down tunneling is allowed only through the
states $|1,1\rangle^{\pm}$ whose weights are small. By similar arguments, one can see that the
discharging process activates two EMSs, namely $|\varphi_{0,1/2}\rangle$ and $|\varphi_{0,-3/2}\rangle$.

Both of these states acquire important weights in the RDO for $E/D\simeq 1/25$ so the average total spin
can no longer be associated to a well defined value of the molecular spin. We therefore conclude that the enhanced QTM
between $Q=1$ states damages the efficiency of the turnstile protocol even if the EMSs involved in transport are not mixed.

Let us note that the possibility to prepare a single EMS is not obvious
as an open system is generally described by a mixed state. Our simulations also show that in the quantum turnstile
regime one controls the transitions between any pair of intermediate molecular states $(|\varphi_{0,m}\rangle, |\varphi_{0,m-1}\rangle)$ along
a pumping cycle, in contrast to the full magnetic switching which involves only the pair
$(|\varphi_{0,S}\rangle, |\varphi_{0,-S}\rangle)$.

Finally, we mention that if the SMM has an integer spin one cannot expect an accurate turnstile operation 
because the transverse anisotropy induces strong mixing between quasidegenerate EMSs (e.g. between 
$|\varphi_{0,1}$ and $|\varphi_{0,-1}$).

\section{Conclusions}

In the present work we address the transient transport regime and turnstile pumping
across a single-molecule magnet coupled to external leads.
The time-dependent evolution of the molecular states has been discussed in detail
and signatures of the electrically induced magnetic switching on the transient currents were predicted. For ferromagnetic leads with antiparallel spin polarizations the turnstile protocol allows the stepwise
writing and reading of excited molecular states.

The evolution of the states along the turnstile operation can be `read' indirectly from the behavior of
the transient currents. More precisely, by recording the charging and discharging currents one can monitor the
evolution of the system and identify the regimes where its density matrix is described by a {\it single} 
empty molecular state. This is somehow in contrast to the situation when the leads are simply normal
metals and where the control of the excited spin states cannot be achieved as the molecular spin
is reversed continuously.

We show that the transverse anisotropy leads to the hybridization of nearly degenerate one-particle states which
subsequently relax to empty molecular states with different values of the total spin. However, this dephasing
effect can be reduced by applying a moderate perpendicular magnetic field.

Another useful application of the turnstile regime that we address here is the possibility to
 mix several excited spin states (viewed as magnetic qubits) during the discharging cycles when
the source electrode is normal and the drain electrode is ferromagnetic. Note that the short
rise time of the switching functions used in our simulations is not essential for
the turnstile operation and slower switching functions could be in principle selected to achieve a better resolution
of the transient peaks. Our predictive simulations clearly emphasize the potential of the molecular
turnstiles as promising candidates for molecular spintronics.
As a method approach we have used the generalized master equation formalism adapted to the turnstile configuration.

\section*{Acknowledgments}
V.M. and I.V.D. acknowledge financial support from PNCDI2 program (grant PN-II-ID-PCE-2011-3-0091) and from grant No.\ 45N/2009.
V.M., I.V.D. and B.T. acknowledge financial support from ANCS-TUBITAK Bilateral Programme COBIL 603/2013 and 112T619.
B.T. also thanks TUBA for support.


\begin{thebibliography}{10}
\bibitem{good}
It can be easily shown that $[H_M,{\hat S}_z^t]=0$ if $E=0$.
\bibitem{Fermi}
For non-interacting leads $\rho_{{\rm el}}$ is defined through the identity
${\rm Tr}_{{\cal F}_L}\{ \rho_{{\rm el}}a^{\dagger}_{q_{\alpha}\sigma}a_{q_{\beta}\sigma}\}=\delta_{\sigma\sigma'}
\delta_{\alpha\beta}f_{\alpha}(\epsilon_{q_{\alpha}})$, where $f_{\alpha}(\epsilon_{q_{\alpha}})$ is the Fermi function
associated to the lead $\alpha$.
\end{thebibliography}
\end{document}